\documentclass[twocolumn,superscriptaddress,prd,showkeys,showpacs,nofootinbib]{revtex4-1}

\usepackage{graphicx}
\usepackage[colorlinks = true,
            linkcolor = cyan,
            urlcolor  = blue,
            citecolor = red,
            anchorcolor = blue]{hyperref}\usepackage{color}
\usepackage{amssymb}
\usepackage[nointegrals]{wasysym}
\usepackage{amsthm}
\usepackage{textcomp}
\usepackage{mathtools}
\usepackage{lineno}

\usepackage{comment}
\usepackage[separate-uncertainty,retain-explicit-plus,per-mode = symbol]{siunitx}
\usepackage{url}

\newcommand{\arsev}{\mbox{$^{37}$Ar}}

\newcommand{\us}{$\mu$s}

\begin{document}

\title{Electron extraction efficiency study for dual-phase xenon dark matter experiments}

\author{J.~Xu} \email[Corresponding author, ] {xu12@llnl.gov}\affiliation{Lawrence Livermore National Laboratory, 7000 East Ave., Livermore, CA 94551, USA}  
\author{S.~Pereverzev}  \affiliation{Lawrence Livermore National Laboratory, 7000 East Ave., Livermore, CA 94551, USA}  
\author{B.~Lenardo} \altaffiliation{Current address: Stanford University, Physics Department, 450 Serra Mall Stanford, CA 94305, USA} \affiliation{Lawrence Livermore National Laboratory, 7000 East Ave., Livermore, CA 94551, USA} \affiliation{University of California Davis, Department of Physics, One Shields Ave., Davis, CA 95616, USA} 
\author{J.~Kingston} \altaffiliation{Current address: The University of Chicago, Division of the Physical Sciences, 5801 South Ellis Avenue Chicago, Illinois 60637, USA} \affiliation{Lawrence Livermore National Laboratory, 7000 East Ave., Livermore, CA 94551, USA}  
\author{D.~Naim} \affiliation{University of California Davis, Department of Physics, One Shields Ave., Davis, CA 95616, USA} 
\author{A.~Bernstein} \affiliation{Lawrence Livermore National Laboratory, 7000 East Ave., Livermore, CA 94551, USA}  
\author{K.~Kazkaz} \affiliation{Lawrence Livermore National Laboratory, 7000 East Ave., Livermore, CA 94551, USA}  
\author{M.~Tripathi} \affiliation{University of California Davis, Department of Physics, One Shields Ave., Davis, CA 95616, USA} 

\date{\today}

\begin{abstract}
Dual-phase xenon detectors are widely used in dark matter direct detection experiments, 
and have demonstrated the highest sensitivities to a variety of dark matter interactions. 
However, a key component of the dual-phase detector technology--the efficiency of charge extraction from liquid xenon into gas--has not been well characterized. 
In this paper, we report a new measurement of the electron extraction efficiency (EEE) in a small xenon detector using two mono-energetic decay features of \arsev. 
By achieving stable operation at very high voltages, 
we measured the EEE values at the highest extraction electric field strength reported to date. 
For the first time, an apparent saturation of the EEE is observed over a large range of electric field; 
between 7.5~kV/cm and 10.4~kV/cm extraction field in the liquid xenon the EEE stays stable at the level of 1\%(kV/cm)$^{-1}$. 
In the context of electron transport models developed for xenon, 
we discuss how the observed saturation may help calibrate this relative EEE measurement to the absolute EEE values. 
In addition, we present the implications of this result not only to current and future xenon-based dark matter searches, 
but also to xenon-based searches for coherent elastic neutrino-nucleus scatters.
\end{abstract}


\maketitle




\section{Introduction}
\label{sec:intro}

Dual-phase xenon time projection chambers (TPCs) have demonstrated exceptionally high sensitivities to various dark matter candidates, 
including both weakly interacting massive particles (WIMPs)~\cite{XENON1T_2018, LUX2016_Run3_4, PandaX2018_EFT} 
and leptophilic dark matter~\cite{XENON2015_Leptophilic, LUX2018_ERMod}. 
They also offer a promising, scalable method for measurement of coherent scattering of reactor antineutrinos~\cite{Hagmann2004_CENNS}. 
Xenon TPCs enjoy high sensitivity to these rare interactions primarily due to their extremely low intrinsic radioactive background levels 
and the efficient collection of both scintillation photons and ionization electrons produced by particle interactions~\cite{Bolozdynya1995_XeTPC, Yamashita2003_XeTPC}. 
The latter feature enables such detectors to obtain accurate three-dimensional position reconstruction capabilities 
and powerful particle type identification~\cite{Yamashita2003_XeTPC, XENON2018_ERNRDist}. 

A typical dual-phase xenon TPC consists of liquid active volume in the bottom 
and a thin gas layer above it. 
Particle interactions in the dense liquid can produce atomic excitations and ionizations~\cite{Doke2002_ArXe}. 
The excitations lead to the emission of prompt scintillation photons. 
The ionization electrons can either recombine with ions 
to produce additional scintillation, 
or escape the recombination process and become quasifree electrons~\cite{Lenardo2015_XeLightCharge}. 
To detect these electrons, 
an electric field is applied to drift them to the liquid surface and extract them into the gas, 
where they produce electroluminescence (EL) light~\cite{Bolozdynya1995_XeTPC}.  
Collection of this secondary EL light gives an amplified measurement of the ionization signal strength, 
which improves the overall energy resolution, 
and enhances position reconstruction and particle identification. 
In addition, the EL amplification mechanism also allows modern xenon TPCs to detect faint signals down to single extracted electrons in the gas with $\mathcal{O}$(100\%) efficiency, 
opening up the possibility of detecting extremely low energy interactions, 
such as that expected from scattering of low-mass WIMPs, certain dark-sector dark matter particles and reactor antineutrinos with xenon~\cite{XENON2011_S2Only, Essig2012_SubGeVXENON10, Essig2017_SubGeVXENON100, Hagmann2004_CENNS}. 
In such cases, the scintillation signal becomes too small to be detectable but a handful of ionization electrons may be collected. 

Although dual-phase xenon TPCs are currently in wide use in dark matter search experiments, 
the efficiency of extracting ionization electrons from liquid xenon into gas has not yet been fully characterized. 
In all existing measurements~\cite{Gushchin1979_EEE, XENON2014_SE, PIXeY2018_EEE}, 
the electron extraction efficiency (EEE) is observed to increase monotonically with the extraction electric field even at the highest probed fields, 
while a saturation should occur if 100\% EEE is approached. 
In addition, measurements by different groups do not fall within one another's experimental uncertainties. 
Part of this discrepancy may be attributed to the different assumptions of full electron extraction in these measurements, 
which can be clarified if 100\% EEE is experimentally measured. 

Furthermore, understanding the EEE is important for the following reasons. 
First, conflicting EEE values result in systematic differences in the ionization energy scale between experiments, 
which directly affects the derived dark matter search sensitivity, 
as well as the derived reactor antineutrino coherent scatter rate. 
Second, incomplete electron extraction is thought to be one of the leading causes of relatively high rates of few-electron events 
observed in low-background xenon TPC experiments~\cite{ZEPLIN2011_SE, XENON2014_SE, LUX2016_SE, SorensenElectronBG_2018}. 
This is a limiting background in searches for WIMPs of $m_\chi < 10~GeV/c^2$ and MeV-scale dark-photon-mediated dark matter interactions, 
which are expected to produce $\mathcal{O}$(1)- to $\mathcal{O}$(10)-electron ionization events in xenon targets~\cite{XENON2011_S2Only, Essig2012_SubGeVXENON10, Essig2017_SubGeVXENON100}. 
Reactor antineutrino experiments using dual-phase noble liquid TPCs are expected to have a very similar signature to that arising from these dark matter models 
- a rapid falling energy spectrum below 10 ionization electrons. 
Current xenon TPC experiments all operate at insufficient extraction electric fields, 
at which electron extraction is incomplete and the EEE values are uncertain. 
Therefore an improved understanding of the EEE can help interpret existing and future data, 
as well as provide a path forward to maximize the scientific reach of future xenon-based experiments. 

In this paper, 
we report a new EEE measurement conducted at the highest extraction electric field up to date, 
providing the most comprehensive EEE calibration for dual-phase xenon TPCs. 
For the first time, a clear saturation of the EEE is observed over a large electric field window, 
suggesting that 100\% EEE may have finally been demonstrated. 

This paper is organized as follows: 
in Section~\ref{sec:mechanism}, we  review theories on how electrons in the liquid may be extracted into gas 
and experimental techniques that have been used to measure the EEE; 
in Section~\ref{sec:detector}, we  describe our EEE measurement method and the apparatus used in this experiment; 
Section~\ref{sec:ana} explains the data analysis techniques; 
Section~\ref{sec:result}  presents the results of this measurement 
and discusses the implications to current and future dark matter searches; 
in Section~\ref{sec:concl}, we summarize the conclusions. 

\section{Electron emission mechanisms and overview of EEE experiments}
\label{sec:mechanism}

\begin{figure}[h!]
\centering
\includegraphics[width=.49\textwidth]{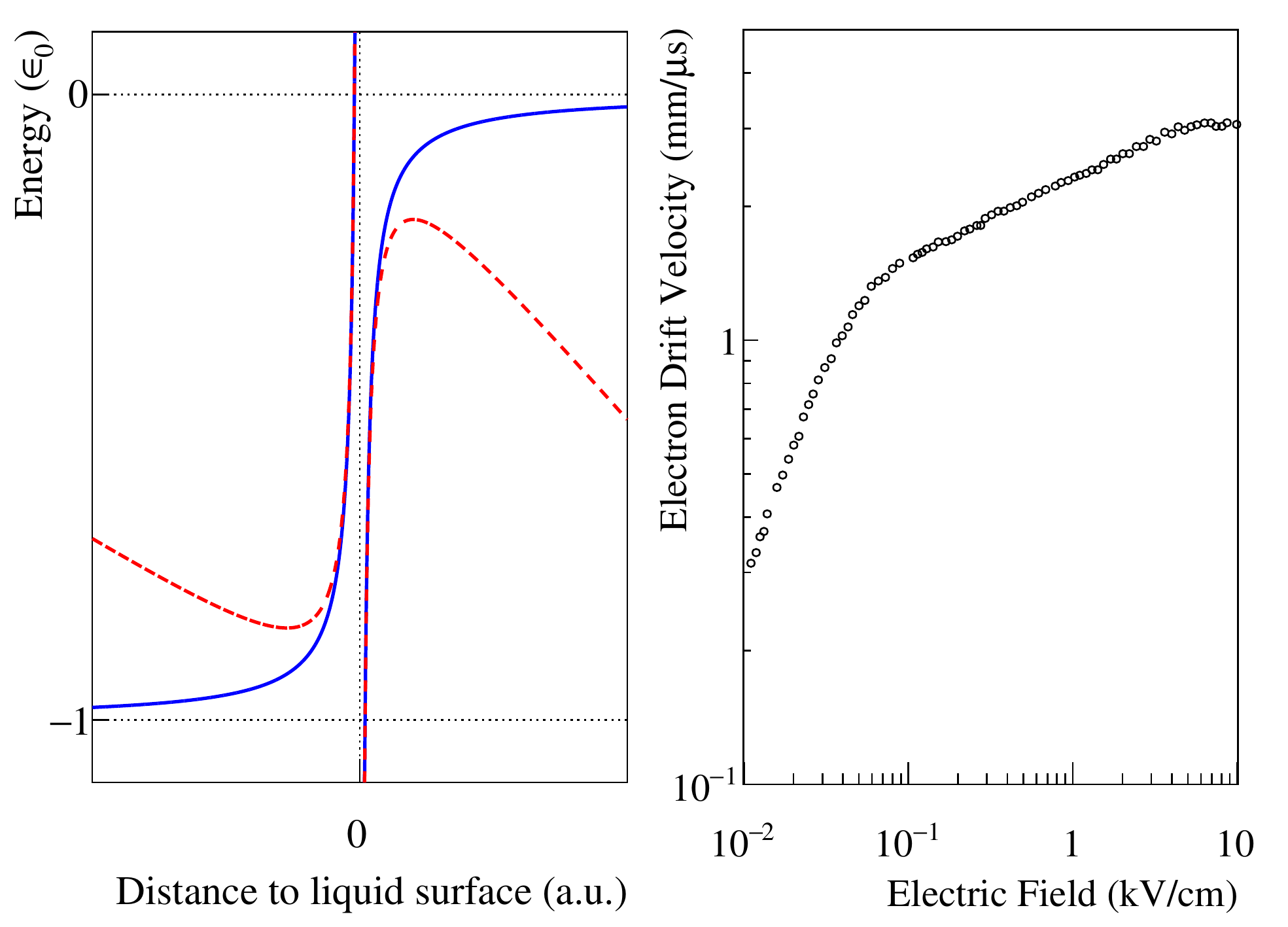}
\caption{{\bf Left:} Illustration of the electronic potential barrier near the liquid-gas boundary (liquid: $x<0$; gas: $x>0$) in a dual-phase xenon TPC, 
  with (dashed line, red) and without (solid line, blue) an external electric field~\cite{Bolozdynya1995_XeTPC}. 
  The energy unit $\epsilon_0$ is the barrier height at the liquid surface without electric fields. 
  {\bf Right:} The measured electron drift velocity in liquid xenon as a function of electric field~\cite{Gushchin1982_ETransport}, 
  illustrating the energy transfer from the electric field to electrons in the liquid. 
  More measurement results can be found in \cite{Miller1968_ETransport} and \cite{EXO2017_EDrift}.}
\label{fig:etransport}
\end{figure}

Due to their large atomic size, 
Xe atoms can be polarized by excess electrons and thereby exert an attractive force on them. 
This effect is stronger in the liquid phase than that in the gas due to the dielectric constant difference. 
As a result, 
the ground-state energy of a quasifree electron is lower in the liquid than that in the gas, 
forming an energy barrier of $\sim$0.6-0.85~eV at the liquid-gas boundary~\cite{Reininger1982_EinXe, Bolozdynya1995_XeTPC, Gushchin1982_EEE}. 
For an electron near the liquid surface, 
the net effect of the inhomogeneous polarizations in the two different dielectric materials 
may be approximated as a mirror charge on the other side of the boundary,  
which adds to the ground potential difference in the medium~\cite{Bolozdynya1995_XeTPC}, 
as illustrated in Figure \ref{fig:etransport} (left, solid line, blue). 
In order for electrons in the liquid to emit into the gas, 
they have to carry a high enough momentum in the direction towards the gas to overcome the energy barrier. 
At liquid xenon temperature ($\sim$175 K), 
the thermal energy of electrons is approximately 0.015~eV, 
too low to produce significant thermal electron emissions. 

When an external electric field is applied to the liquid, 
quasifree electrons will drift along the direction of the field, 
and gain energy from electric field acceleration between collisions.  
Figure \ref{fig:etransport} (right) shows the increasing drift velocity of electrons in liquid xenon with electric field~\cite{Gushchin1982_ETransport}, 
demonstrating the energy transfer from the static electric field to the electrons. 
The dynamics of electrons in noble gases/liquids under an external electric field has been extensively modeled 
in condensed matter physics~\cite{Frost1964_ETransport, CohenLekner1967_ETransport, Atrazhev1981_HotE, Gushchin1982_ETransport, Atrazhev1985_EHeating, Boyle2016_ETransport, Gordon2001_ETransport, Sakai1984_ETransport}. 
A notable approach in these studies is the Cohen-Lekner transport theory, 
which introduces a structure factor $S(\vec{K})$ to approximate the collective scattering of hot electrons with atoms in condensed phases, 
and solves the modified Boltzmann equation to derive the electron energy distribution function~\cite{CohenLekner1967_ETransport}. 
Implementations of this method have successfully reproduced experimental data on the field dependence of 
electron mobility and diffusion properties in liquid xenon~\cite{Gushchin1982_ETransport, Atrazhev1985_EHeating, Boyle2016_ETransport, Sakai1984_ETransport}. 
They also predict that the electron energy will increase monotonically with electric field 
until it becomes comparable to the first excitation energy level ($\sim$8~eV for liquid xenon)~\cite{Gushchin1982_ETransport, Atrazhev1985_EHeating}. 
The increasing electron energy will then enable electron emission from the liquid into the gas, 
which has an experimentally observed threshold of $\sim$1.5-2 kV/cm, 
with the extraction efficiencies increasing with applied field strength~\cite{Gushchin1979_EEE, XENON2014_SE, PIXeY2018_EEE}. 

In addition to increasing the electron energy, 
an electric field at the liquid-gas boundary also reduces the barrier height and width (Schottky effect)~\cite{Bolozdynya1995_XeTPC}, 
as illustrated in Figure \ref{fig:etransport} (left, dashed line, red). 
These effects will make it easier for electrons in the liquid to be emitted into the gas, 
and also reduces the chance for an emitted electron to be reflected back into the liquid~\cite{Gushchin1982_EEE}. 
It is worth noting that when a cloud of electrons approach the liquid surface, 
although only electrons with high momentum values in the direction towards the gas are emitted in the first place, 
the rest of electrons with sufficient kinetic energy can gain the ``correct'' momentum after scattering with xenon atoms. 
Thanks to the much smaller electron mass compared to that of the xenon atoms, 
electrons barely lose any energy during elastic scattering with xenon; 
it has been estimated that electrons approximately lose 0.1\% of their kinetic energy for their directions to be flipped~\cite{Gushchin1982_EEE}. 
As a result, all electrons with sufficient kinetic energy in pure liquid xenon should be able to emit into the gas after several attempts within nanoseconds. 

A few experimental efforts have been made to directly evaluate the EEE dependence on electric field~\cite{Gushchin1979_EEE, XENON2014_SE, PIXeY2018_EEE}. 
An early measurement by Gushchin {\it et al.} attempted to derive the absolute EEE values 
by comparing the ionization current collected by an electrode 
when it was above liquid xenon and when it was immersed in the liquid~\cite{Gushchin1979_EEE}. 
The result suggested that electron extraction starts below 2~kV/cm and the extraction efficiency keeps increasing up to the highest attained field of 4.3~kV/cm. 
Recent experiments took a different approach by measuring the number of extracted electrons from mono-energetic decays 
at different extraction fields in dual-phase xenon TPCs~\cite{XENON2014_SE, PIXeY2018_EEE}. 
Given that the average number of electrons available for extraction is constant for monoenergetic sources, 
the numbers of extracted electrons are proportional to the EEE values at the liquid-gas boundary. 
However, without independently measuring the total numbers of electrons produced by the sources, 
EEE values measured in these experiments may carry an unknown scaling factor and are usually referred to as relative measurements. 
As explained in Section~\ref{sec:result}, 
a definitive saturation in the relative EEE value is a strong indication of approaching 100\% absolute EEE. 
Hint of a possible saturation was suggested by XENON100 in a narrow field window of $\sim$0.5~kV/cm at $>$5~kV/cm~\cite{XENON2014_SE}, 
but a more recent measurement argued that the EEE still increases in this field range~\cite{PIXeY2018_EEE}. 

An indirect method to obtain absolute EEE values is to use the anticorrelation 
between the charge and scintillation channels in liquid xenon~\cite{Aprile2007_XeScintIon}.  
For mono-energetic radioactive sources, the anti-correlation can be formulated as: 
\begin{equation}
\label{eq:corr}
\frac{N_{\gamma, ob}}{\epsilon_{\gamma}} + \frac{N_{e, ob}}{\epsilon_{e}} = N_q(E) 
\end{equation}
where $N_{\gamma, ob}$ and $N_{e, ob}$ are the detected numbers of scintillation photons and electrons, 
$\epsilon_{\gamma}$ and $\epsilon_{e}$ are the experimental efficiencies, 
and $N_q$ is the total number of quanta produced by the source 
and is often modeled as a linear function of energy $N_q=E/W$~\cite{DahlThesis, LUX2018_Run3PRD}. 
Given direct knowledge of $\epsilon_{\gamma}$ or $N_q$, 
one can derive the absolute EEE ($\epsilon_{e}$) value using the measured anticorrelation between $N_{\gamma, ob}$ and $N_{e, ob}$. 
This method has been attempted by both LUX~\cite{LUX2016_Run3_4} and XENON1T~\cite{XENON1T2017_Experiment}, 
but the obtained EEE values are subject to uncertainties in the $N_q$. 

\section{Apparatus and measurements}
\label{sec:detector}

\begin{figure}[h!]
\centering
\includegraphics[width=.45\textwidth]{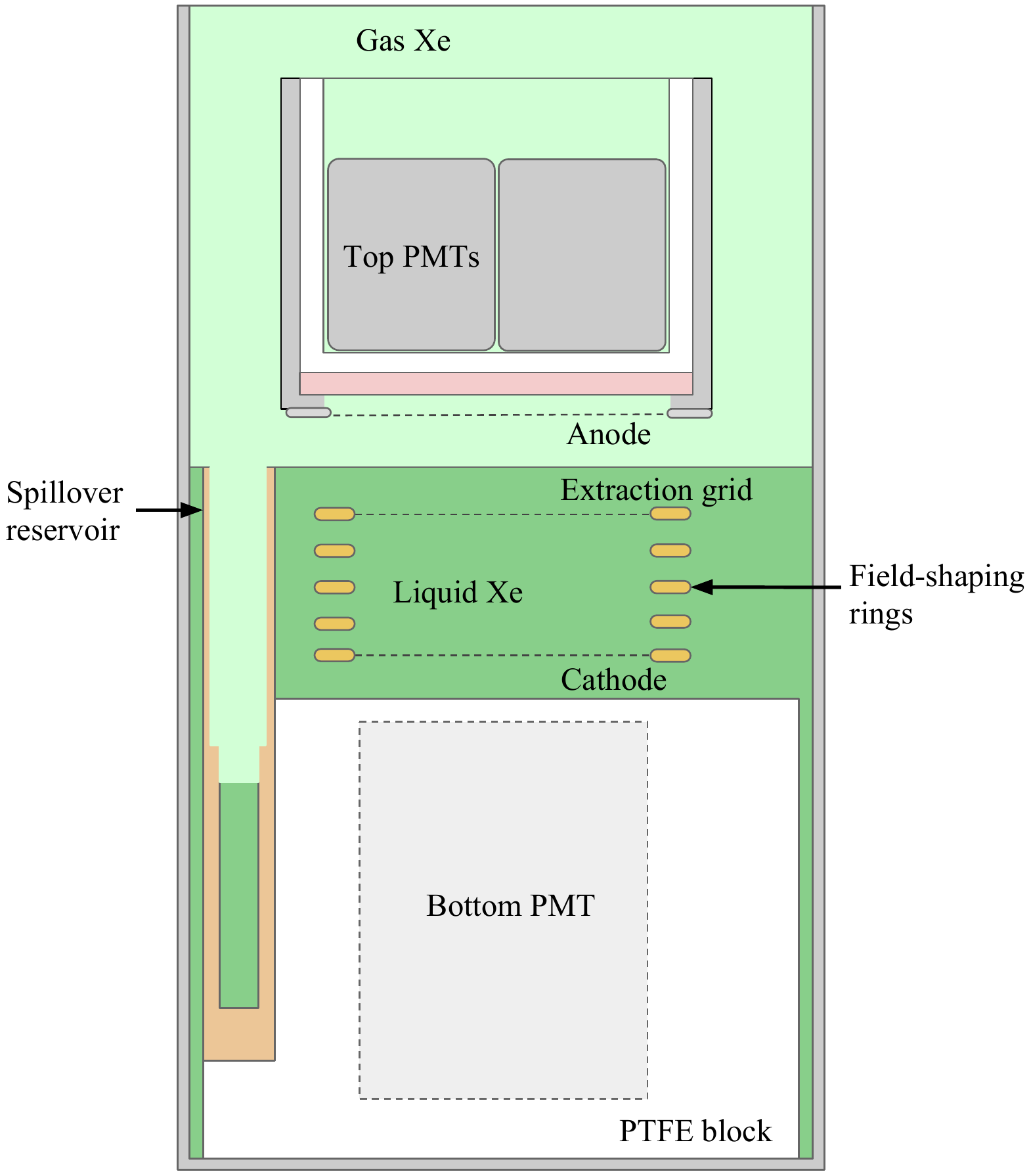}
\caption{Schematic of the dual-phase xenon detector. 
The active xenon volume of $\diameter$5~cm$\times$2.5~cm is contained in an electric field-shaping cage 
made of copper rings and stainless steel grids connected through Giga-Ohm resistors. 
vacuum-ultraviolet (VUV)-sensitive photomultipliers (PMTs) collect the xenon scintillation and EL light from both above and below the active volume. }
\label{fig:detector_schematic}
\end{figure}

The measurements presented in this work were made with the XeNu apparatus, 
a prototype dual-phase TPC optimized for measuring low-energy ionization signals. 
The detector consists of a cylindrical active volume containing approximately 140~g of liquid xenon in an electric field-shaping cage. 
Electrons from ionization events in this active volume are drifted upwards via applied electric fields 
and extracted into a gas region. 
The EL light produced by extracted electrons in the gas is detected by an array of VUV-sensitive PMTs: 
four Hamamatsu R8520-406 PMTs (2.6~cm$\times$2.6~cm) measure the light from above, giving horizontal position information, 
and one Hamamatsu R8778 PMT ($\diameter$5.6~cm) is immersed in the liquid below the active xenon to increase the light collection efficiency. 
To enhance position reconstruction accuracy and avoid possible high voltage instabilities, no reflector was installed around the electric field cage. 
A schematic of the detector internals is shown in Fig.~\ref{fig:detector_schematic}.

Three grids supply the electric fields for drifting and extracting electrons: 
the cathode, extraction grid, and anode. 
One additional grounded grid shields the bottom PMT from the high voltage applied to the cathode.
All grids are made from a stainless steel hexagonal mesh (acid etched, 90\% transparent at normal incidence) press-fit into metal rings for mechanical stability~\footnote{The first measurement reported in this work used a stainless steel mesh soldered onto a Cirlex ring for the anode.}. 
The active volume is surrounded by three additional copper rings which step down in voltage to
create a uniform electric field throughout. 
The liquid level in the detector is set to be approximately in the middle between the extraction grid and the anode, 
and its exact position is determined by the height of a spillover reservoir. 
High voltage (HV) to the extraction grid and the cathode was provided through two custom-made HV feedthroughs. 
During data taking, up to 21~kV HV was applied to the extraction grid and up to 22~kV was applied to the cathode. 
The anode was grounded in all measurements. 

Xenon is circulated continuously through a closed loop during detector operation. Liquid is drawn from the spillover reservoir into
a dual-phase heat exchanger, where it is evaporated and pumped through a SAES MonoTorr getter for gas-phase purification. The purified gas is
then cooled down by the heat exchanger, re-condensed and delivered directly to the bottom of the detector to avoid disturbing
the liquid surface. The circulation rate is regulated by an MKS mass flow controller (MFC) and operated between 1.2 and 2.1 Standard Liter Per Minute (SLPM), 
turning over all $\sim$1.5~kg of xenon in the system in 2~-~4~hours, ensuring sufficient xenon purity for the experiment. 

To provide a source of low-energy, mono-energetic signals in our detector for the EEE studies, 
\arsev\ gas was injected directly into the xenon circulation loop. 
This technique has been used previously to calibrate both xenon \cite{Boulton2017_Ar37,Akimov2014_Ar37} and argon \cite{Sangiorgio2013_Ar37} detectors. 
The source decays via electron capture, 
with captures from either the L-shell or K-shell depositing a total of 0.27~keV or 2.8~keV, respectively, via a cascade
of X-rays and/or Auger electrons. The L/K capture branching ratio is $\sim$0.1 \cite{SantosOcampo1960_Ar37}.
For this work, the \arsev\ calibration source was generated by a 40~hr irradiation of
natural argon gas at the McClellan Nuclear Research Center TRIGA reactor.
Thermal neutron capture on $^{36}$Ar and $^{40}$Ar created the radioisotopes \arsev\ and $^{41}$Ar.
After irradiation, the gas was subject to a two day cool-off period, during which the $^{41}$Ar
($t_{1/2} = 1.8$ hrs) decayed away while \arsev\ ($t_{1/2}=35$ days) remained. 
The remaining gas contained $\sim$100~$\mu$Ci of \arsev\ dissolved in $\sim$40~g of $^{\text{nat}}$Ar. 
For each measurement, we injected $\mathcal{O}$(10)~Bq of \arsev\ into the detector 
to produce low-energy decays uniformly distributed throughout the liquid active volume.
To suppress ambient gamma-ray backgrounds in the \arsev\ measurements, 
the detector was surround on all sides with at least 5~mm of lead shielding. 

The data analyzed in this work were taken in two separate configurations, 
changing the heights of the liquid and gas volumes between the extraction grid and the anode. 
In the first configuration, the top of the spillover reservoir was set 7.5$\pm$0.5~mm above the extraction grid and 8.4$\pm$0.5~mm below the anode grid. 
In the second, the top of the spillover reservoir was set 8.0$\pm$0.5~mm above the extraction grid and 6.4$\pm$0.5~mm below the anode grid. 
The liquid level was estimated to be 0.5$\pm$0.5~mm above the top of the spillover reservoir due to fluid dynamics~\cite{Pfister2013_LiquidLevel}. 
Because of the smaller distance between the extraction and anode grids in the second configuration, a higher electric field was achieved for the same applied voltage. 
For this reason, the first configuration is referred to as the low field (LF) configuration, 
and the second is referred to as the high field (HF) configuration. 
In addition to the change in grid distance, 
the HF configuration also introduced a 0.5~mm thick PEEK cover on the extraction grid holder in hope of enhancing high voltage stability. 
The effects of charge buildup on the PEEK cover or the Cirlex ring was modeled using a COMSOL electrostatics simulation 
    and was found to have a negligible impact ($<$1\% on the extraction field) on these measurements. 
An extraction electric field of 8.6~kV/cm in liquid xenon was achieved in the LF run, 
and 10.4~kV/cm was achieved in the HF run. 
In the following discussions, 
the electric field value in the liquid above the extraction grid is used unless specified otherwise.
Data from both the LF and HF experiments are presented below.

\section{Data analysis}
\label{sec:ana}

The raw data consisted of digitized waveforms from all five PMTs ($\times$10 amplified linearly) using Struck SIS3316 digitizers (14 bit, 250~MHz). 
For each of the PMT waveforms, 
an adaptive baseline was first calculated to separate PMT output pulses from electronic noise. 
The gain of each PMT was calibrated by evaluating the average size of single photoelectron (p.e.) pulses in regions of low pulse frequency.
PMT saturation effects were observed for pulses with amplitudes of $\sim$1~V after $\times$10 amplification; 
as a result, pulses within 5~ms following large pulses were excluded from the calibration to avoid biases. 
The gain-normalized waveforms from all the PMTs were then added together to form summed waveforms, 
which were further processed for pulse identification and pulse quantity evaluation.  

In this analysis, a pulse is registered if the integral of the summed waveform within a preset window (depending on extraction field) 
exceed an adjustable threshold, 
which is set to be below half of the value of the single electron integral for this dataset. 
Piled-up pulses that do not cross the absolute threshold between peaks are also resolved 
by comparing the peak-to-valley difference with a relative threshold. 
For every pulse identified, 
we calculated the pulse area $A$ in the unit of p.e. 
and several other characterization parameters including the pulse integral times and top bottom asymmetry (TBA). 
The pulse integral time  $t_f$ is defined as the time when the pulse integral rises to a specified percentage of the overall pulse area, 
relative to the start time of the pulse. 
In the analysis, we usually use the difference between two $t_f$'s for pulse identifications. 
For example, $W_{10-50}=t_{50}-t_{10}$ is a measure of the pulse width between 10\% pulse integral and 50\% pulse integral. 
A typical liquid xenon scintillation pulse has a $W_{10-50}$ value of $<$ 0.1 \us, 
while that of a typical EL pulse is at the order of 0.5-3 \us. 
The TBA parameter is defined as $(A_T-A_B)/(A_T+A_B)$, 
where $A_T$ and $A_B$ are the total pulse area in the top PMTs and in the bottom PMT, respectively. 
The TBA is usually close to -0.8 for scintillation pulses in the liquid and 0.4-0.5 for EL pulses in the gas. 
Examples of the $W_{10-50}$ and TBA distributions can be found in Section~\ref{sec:ar37}. 
For EL pulses, 
the lateral positions of an event were also calculated based on the EL light distribution in the top PMT array, 
using a simple center of gravity (CoG) method. 
The CoG positions were approximately calibrated by comparing the density distribution of \arsev\ decay events 
in the CoG parameter space to that in the real world space (assumed uniform). 

\subsection{Single electron calibration}
\label{sec:se}

\begin{figure}[h!]
\centering
\includegraphics[width=.45\textwidth]{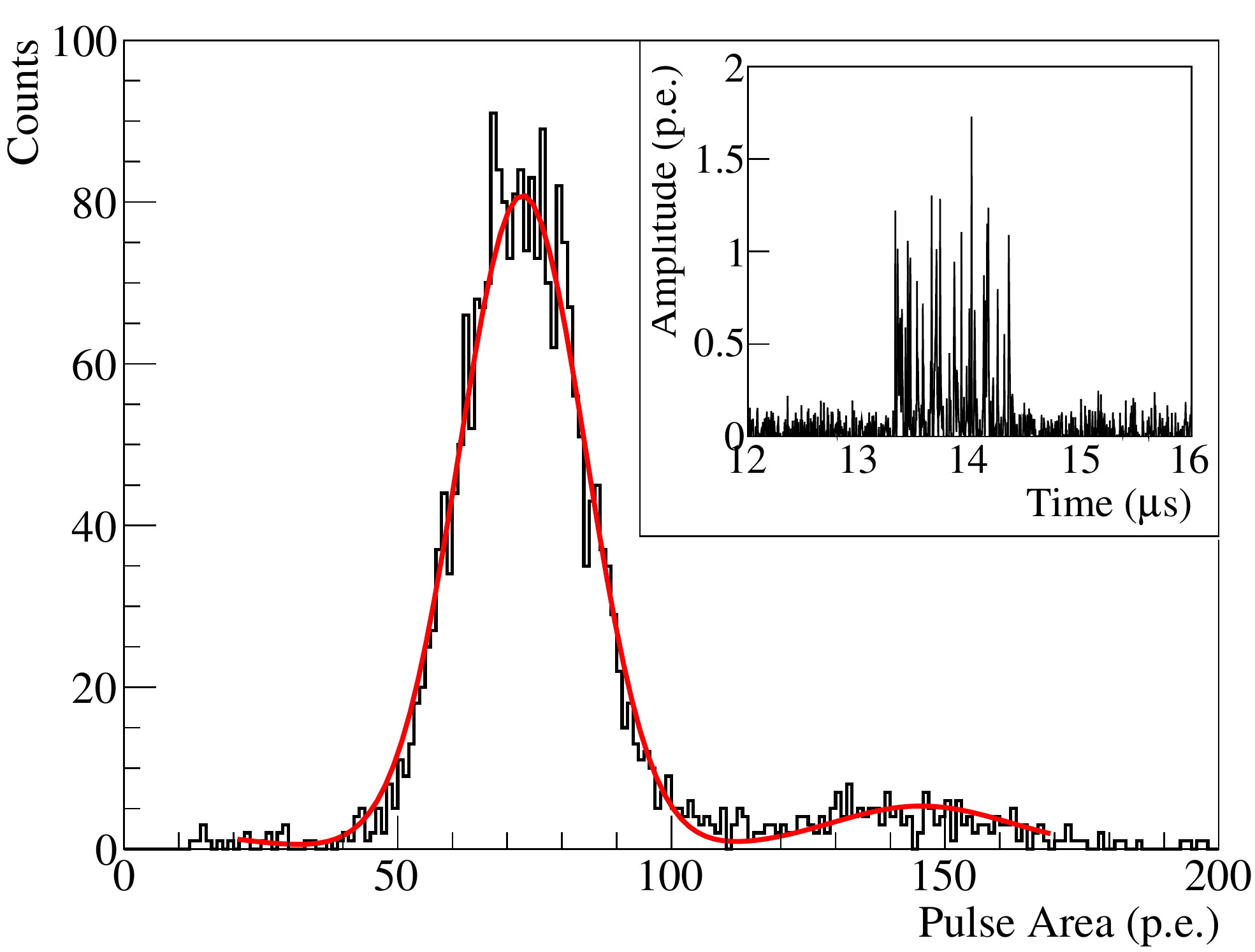}
\caption{The measured SE spectrum at a liquid electric field of 6.0~kV/cm (11.2~kV/cm in the gas) in the LF run. 
The peak around 72~p.e. is from SE pulses and that around 145~p.e. is from double electron (DE) pulses. 
In the fit function, the DE peak position is fixed at twice of that for the SE peak, 
and the width is fixed at $\sqrt{2}$ times of the SE width. 
The inset figure shows a SE waveform in the same dataset. }
\label{fig:se}
\end{figure}

To calculate the number of detected electrons for \arsev\ pulses, 
the average size of SE pulses in the number of p.e. needs to be determined first. 
In this experiment, 
SEs are often observed shortly after multi-electron EL pulses, 
at an estimated total rate of a few tens of Hz, 
offering a convenient way for the SE gain to be calibrated. 
Over the range of electric fields applied to the EL gas region in the \arsev\ measurements, 
the observed SEs contain 30~-~100~p.e. within time windows of $\sim$0.5-3~\us, 
making them relatively easy to identify. 
One example SE pulse is shown in the inset of Figure~\ref{fig:se}, 
which consists of approximately 70~p.e. from all 5 PMTs. 

To evaluate the average SE pulse area at a certain amplification electric field, 
we selected the well-separated SEs that were $>$5 ms after large EL pulses that may have saturated the PMTs. 
Pulses that occurred at or before the trigger time in each event window were also excluded 
to avoid possible biases due to trigger efficiency loss for small pulses. 
In addition, it was required that the pulse width parameter $W_{25-75}$ 
and the TBA of a candidate pulse to not deviate from the mean values for this dataset for more than 1.5$\sigma$ 
to reject misidentified SE pulses and abnormal events. 
Figure~\ref{fig:se} shows the SE spectrum for data acquired at an electric field of 6.0~kV/cm in the LF run, 
where, in addition to a SE peak, a double electron (DE) peak is also observed at approximately two times of the SE peak position. 
Due to the weaker amplification field at the perimeter of the grids and the position dependence of the EL light collection efficiency, 
the observed SE size exhibited a mild dependence on the event position. 
In this analysis only SE pulses emerging in the central 1~cm$\times$1~cm active volume were selected; 
although the EL gain was verified to be approximately uniform at larger radii using \arsev\ data, 
SE events outside this central volume began to get contaminated by events near the perimeter of the TPC due to the limited position resolution of SE pulses. 

\begin{figure}[h!]
\centering
\includegraphics[width=.42\textwidth]{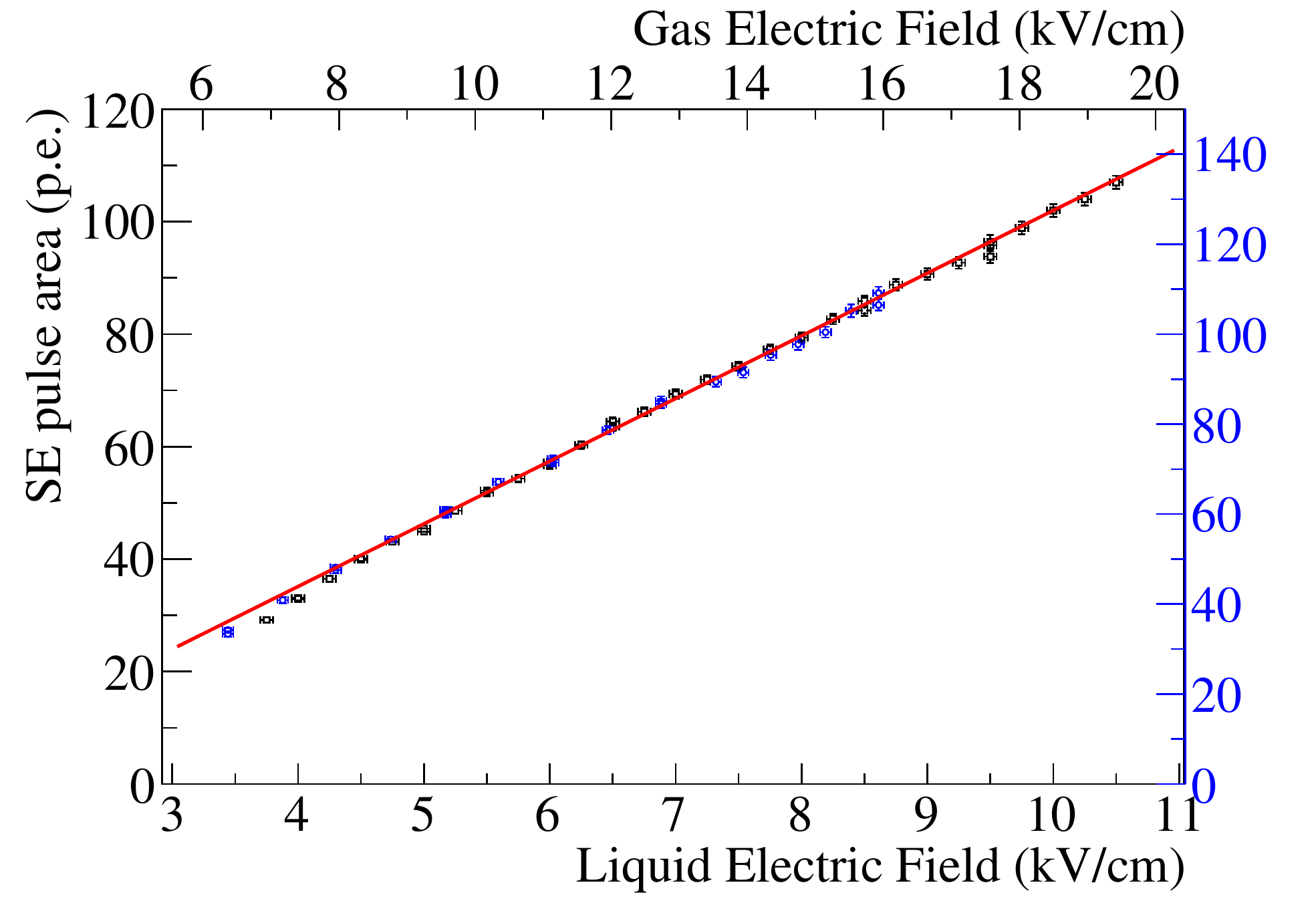}
\caption{The measured SE size as a function of the electric fields applied to the liquid xenon (axis on the bottom) and to the gas (axis on the top) 
  for both the LF data (blue circles, axis on the right) and the HF data (black squares, axis on the left). 
  The error bars on the SE area include the statistical uncertainties from the fit ($\sim$1\%), 
  and that from varying the SE selection cuts by 1$\sigma$ (use an average value of 1\%). 
  The red line shows a linear function fitted to both datasets above 4~kV/cm, 
  after the different EL gap sizes are taken into consideration. 
  The EL threshold is estimated to be 1.6~kV/cm amplification field in the gas. 
}
\label{fig:sevse}
\end{figure}

The dependence of the measured SE gain as a function of the electric fields applied to the liquid and that to the gas, 
which differ by the dielectric constant of the liquid (assumed to be 1.85), 
is shown in Figure~\ref{fig:sevse} for both the LF and HF measurements. 
The electric field value was calculated using a parallel plate approximation, 
with an additional correction based on comparison with COMSOL simulations at the extreme voltage values. 
While Ref.~\cite{PIXeY2018_EEE} reported a large electric field difference of 12\% between COMSOL results and a parallel-plate estimation, 
we calculated a smaller difference of $\sim3\pm1$\%. 
This smaller discrepancy is likely to due to our use of hexagonal grid design compared to the parallel wire design in \cite{PIXeY2018_EEE}.
The SE gain is approximately linear with the EL amplification field and the EL gas gap size, 
similar to the results reported in \cite{PIXeY2018_EEE} and \cite{Monteiro2007_XeScint}. 
Below a liquid extraction field of 4~kV/cm, 
the low detected photon numbers in the SE pulses resulted a relatively large spread in the reconstructed positions, 
causing a non-negligible contamination of large-radii SE pulses in the selected fiducial volume 
and lowering the apparent SE area. 
In the following analysis, 
a linear extrapolation is used to estimate the SE area below 4~kV/cm extraction field. 

\subsection{\arsev\ analysis}
\label{sec:ar37}

Thanks to the small active mass of the xenon TPC and the lead shielding, 
the rate of background events (excluding SEs) in the detector was measured to be less than 10~Hz prior to \arsev\ injection, 
and most of these background occurred in high energy regions above 100~keV. 
On the other hand, 
by controlling the amount of \arsev\ gas injected into the xenon TPC, 
we obtained \arsev\ decay rates of $\mathcal{O}$(10)~Hz in both measurement campaigns. 
Therefore, the rate of ambient radioactive background was negligible in the energy region relevant to this \arsev\ analysis ($<$3~keV). 

For the EEE evaluation, 
only the \arsev\ events in the center of the active volume were selected. 
Owing to the much higher EL signal strength of \arsev\ events than that of SEs, 
their position resolution was greatly improved, 
allowing a larger fiducial cut of 1.8~cm$\times$1.8~cm to be used.
The electric field in this volume was confirmed to be uniform with COMSOL simulations, 
and the observed \arsev\ peak value did not vary with event positions above 5\% level. 
However, as reported by the authors of \cite{MIX2015_Detctor}, 
a position reconstruction algorithm based on light distribution in a four-PMT layout, 
as that used in our detector, 
can misconstruct events on the edge of the detector to be in the center. 
This ambiguity occurs because all PMTs extend comparably small solid angles to the detector perimeters,  
and detected approximately the same number of photons for the edge events, 
mimicking that of lower energy events near the center of the detector.

\begin{figure}[h!]
\centering
\includegraphics[width=.48\textwidth]{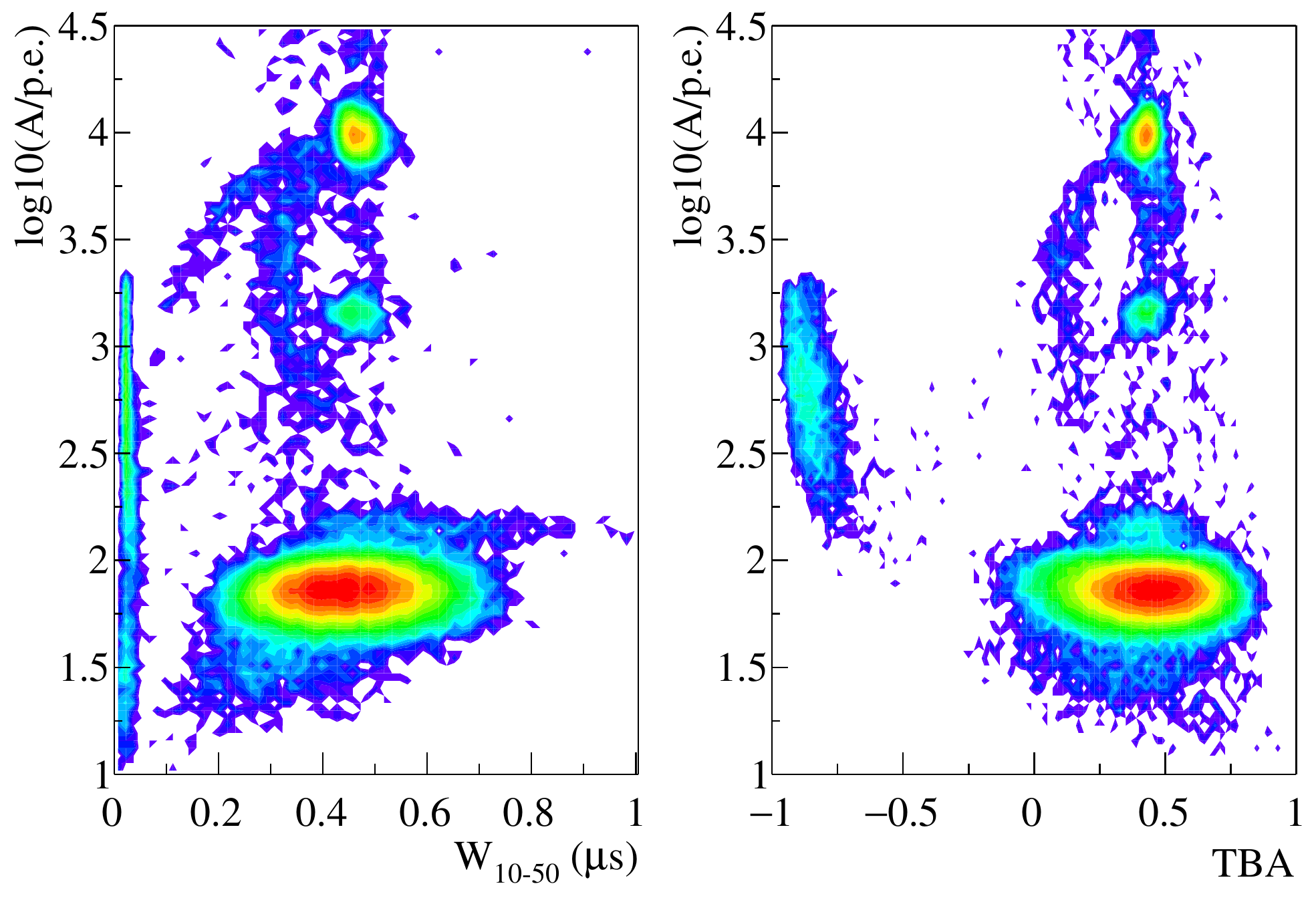}
\caption{The distribution of $W_{10-50}$ (left) and TBA (right) 
as a function of EL pulse area ($A$) for \arsev\ data taken at an extraction electric field of 6.0~kV/cm. 
The events with low $W_{10-50}$ and TBA values are liquid xenon scintillations, 
and the ones with higher values are EL pulses. 
Because only the largest pulse in an event is used for this plot, 
scintillation pulses with areas smaller than SEs are suppressed. 
\arsev\ events in the center of the detector can be observed with high statistics 
around 10$^4$~p.e. (2.8~keV) and 10$^3$~p.e. (0.27~keV) pulse area, 
and the cluster below 100~p.e. are SEs. }
\label{fig:psd}
\end{figure}

These edge events are rejected using the pulse width parameter $W_{10-50}$ and the TBA parameter. 
Figure~\ref{fig:psd} shows the distributions of the $W_{10-50}$ parameter 
and TBA for pulses of different sizes in the \arsev\ data taken at an extraction field of 6.0~kV/cm.  
In this dataset, the events with pulse areas around 10$^4$~p.e. are determined to be the K-shell \arsev\ decays, 
and the events around 10$^3$~p.e. are from the L-shell \arsev\ decays, 
while the ones around 70~p.e. are SEs. 
Due to the weaker amplification field and the incomplete EL light collection, 
the edge events tend to have smaller pulse width values, 
low photon numbers, and lower TBA distributions compared to events in the center of the TPC. 
In the following analysis, 
the events within 2 $\sigma$ from the peak $W_{10-50}$ and TBA values are selected. 
The position distributions of the \arsev\ events passing this cut were approximately uniform, 
confirming that the edge events have been sufficiently suppressed. 

\begin{figure}[h!]
\centering
\includegraphics[width=.49\textwidth]{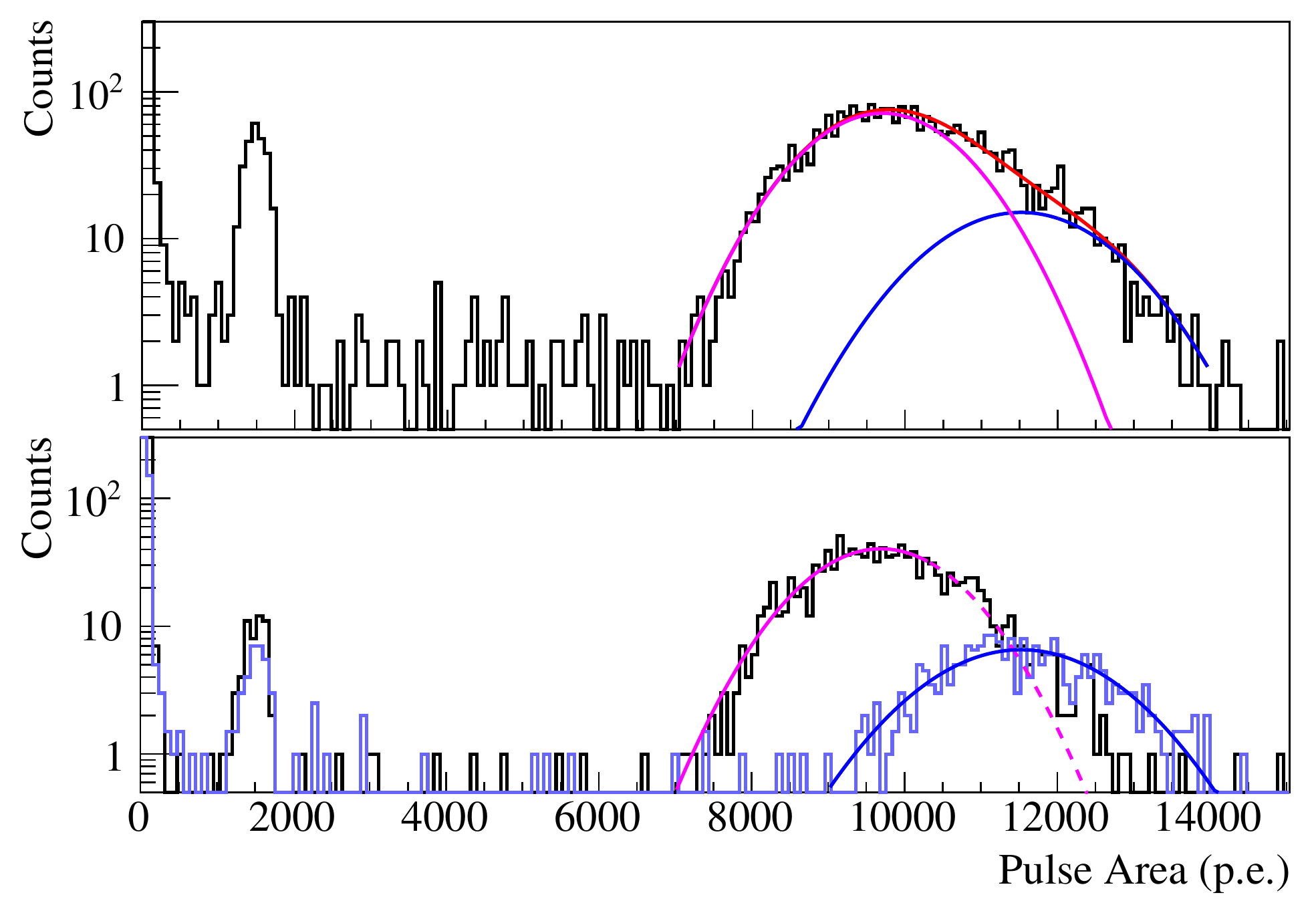}
\caption{Energey spectra of \arsev\ events measured at 6.0~kV/cm extraction field in the LF run. 
{\bf Top: } Spectrum of all selected \arsev\ events in the center of the TPC. 
The fit function (red) to the K-shell peak consists a component above the extraction grid, or EG, (blue) and another component below the EG (magenta). 
{\bf Bottom: } Spectrum of \arsev\ events below the EG (black), selected using a drift time cut, 
and above the EG (light blue), measured with reversed drift electric field. 
Solid lines are the corresponding Gaussian fits, 
and the dashed line shows the function values outside the fit region. }
\label{fig:arspect}
\end{figure}

Figure \ref{fig:arspect} (top) shows the \arsev\ energy spectrum for the 6.0~kV/cm LF data 
with additional $W_{10-50}$ and TBA cuts applied. 
Both the K-shell (2.8~keV) and L-shell (0.27~keV) \arsev\ decay features can be clearly identified. 
The K-shell \arsev\ peak, however, 
exhibits an asymmetric shape with a enhanced tail on the high energy side, 
which is attributed to the contribution of \arsev\ decay events above the extraction grid (EG). 
The electric field in the liquid above the EG was approximately an order of magnitude stronger than that below the EG (nominally at 400~V/cm), 
and as a result, the electron-ion recombination probability in this region was strongly suppressed, 
effectively producing more quasifree electrons that could be drifted, extracted, and collected. 
In addition, as the electron extraction field was changed in the experiment, 
the extent of recombination suppression in this above-EG region also changed, 
resulting a field-dependent background in the EEE study. 
For the lower energy L-shell \arsev\ decays, this effect was not observed, 
possibly because the recombination probability was negligible at all fields due to the extremely low ionization densities, 
as predicted by first-principal simulations in liquid argon~\cite{Foxe2015_LowESim}.  

\begin{figure}[h!]
\centering
\includegraphics[width=.49\textwidth]{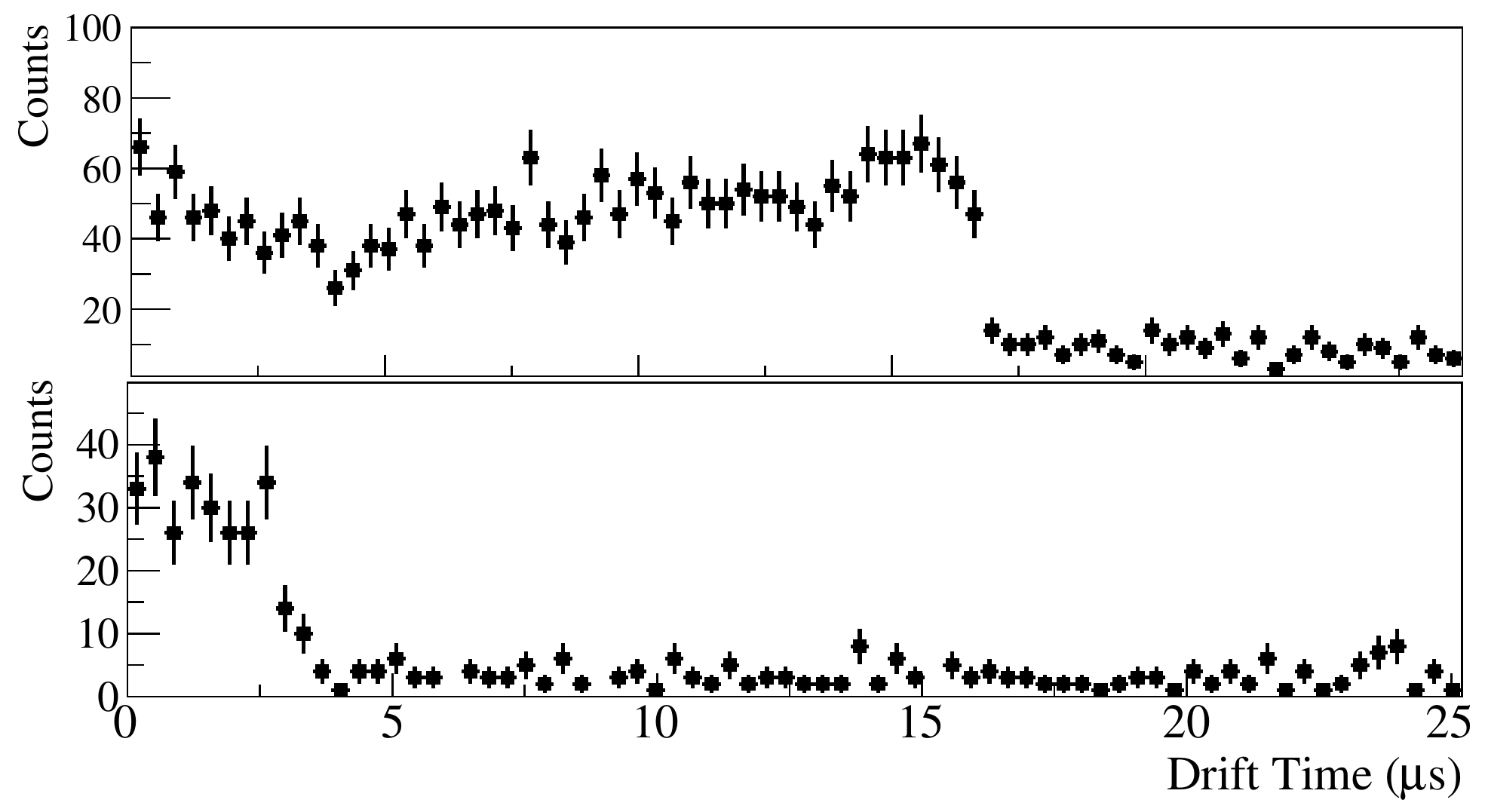}
\caption{Drift time distributions for K-shell \arsev\ decays in the 6.0~kV/cm LF data, 
with nominal drift field (top) 
and with reversed drift field (bottom). 
In the reversed field data, only ionization events above the EG can be detected. }
\label{fig:dt}
\end{figure}

Two different methods were investigated to suppress this above-EG background for the K-shell \arsev\ analysis. 
The first method was to pair EL pulses with prompt scintillations, 
and to use the time separation between the two signals -- usually referred to as the electron drift time -- 
to estimate the depth of the decay site. 
Approximately 60\% of the K-shell \arsev\ decays were observed with a prompt scintillation signal at or above the single p.e. level, 
and thus allow the electron drift time to be calculated. 
Figure~\ref{fig:dt} (top) shows the drift time distribution of K-shell \arsev\ decays in the 6.0~kV/cm LF data, 
where a clear cutoff is observed at 17 \us, 
corresponding to the maximum drift time in the TPC. 
Due to the small size of the \arsev\ scintillation signals, 
random coincidence of single p.e. pulses and EL pulses was non-negligible, 
as illustrated by the events with large drift time values in Figure~\ref{fig:dt} (top).  
It was estimated that $\sim$20\% of the events with normal drift time value were from random coincidence, 
approximately a quarter of which ($\sim$5\%) were from above the EG based on liquid volume considerations ($\sim$7-8~mm liquid above the EG and $\sim$26~mm below the EG). 
The energy spectrum of below-EG K-shell \arsev\ events, selected using a drift time cut, 
is shown in Figure~\ref{fig:arspect} (bottom, black line). 
The L-shell peak in this spectrum is mostly from random coincidence, 
and therefore includes events both above and below the EG. 
For the peak positions of K-shell \arsev\ decays to be determined, 
the high energy tail was excluded from the Gaussian fits to further reduce potential biases from residual above-EG events. 

The second method was to statistically subtract the above-EG \arsev\ decay events. 
In both the LF and HF measurement campaigns, 
we took reversed-drift \arsev\ data 
in which a drift field of $\sim$-400 or $\sim$-800~V/cm was applied to the liquid xenon volume below the EG, 
so only the above-EG \arsev\ decay events could be detected in the gas phase. 
The drift time distribution for the reversed drift field dataset at 6.0~kV/cm extraction field is shown in Figure~\ref{fig:dt} (bottom), 
where the observed events concentrate in regions of drift time $<$3~\us, 
as expected from drift time estimation.  
The \arsev\ spectrum obtained from this dataset is shown in Figure~\ref{fig:arspect} (bottom, light blue histogram): 
this K-shell \arsev\ peak is observed above that of the below-EG events; 
the peak shape is symmetric and can be fit well with a simple Gaussian. 
The peak position and spread of the above-EG \arsev\ events was then used to constrain 
the fit for the below-EG \arsev\ event spectrum. 
Figure~\ref{fig:arspect} (top) includes a double-Gaussian fit to the K-shell \arsev\ spectrum with the aforementioned constraints, 
and the fit result is in excellent agreement with that evaluated with the drift time cut. 
Because of the larger uncertainties in the double-Gaussian fits than that from the drift time analysis, 
only the results from the drift time method is used for the rest of this analysis. 

\begin{figure}[h!]
\centering
\includegraphics[width=.45\textwidth]{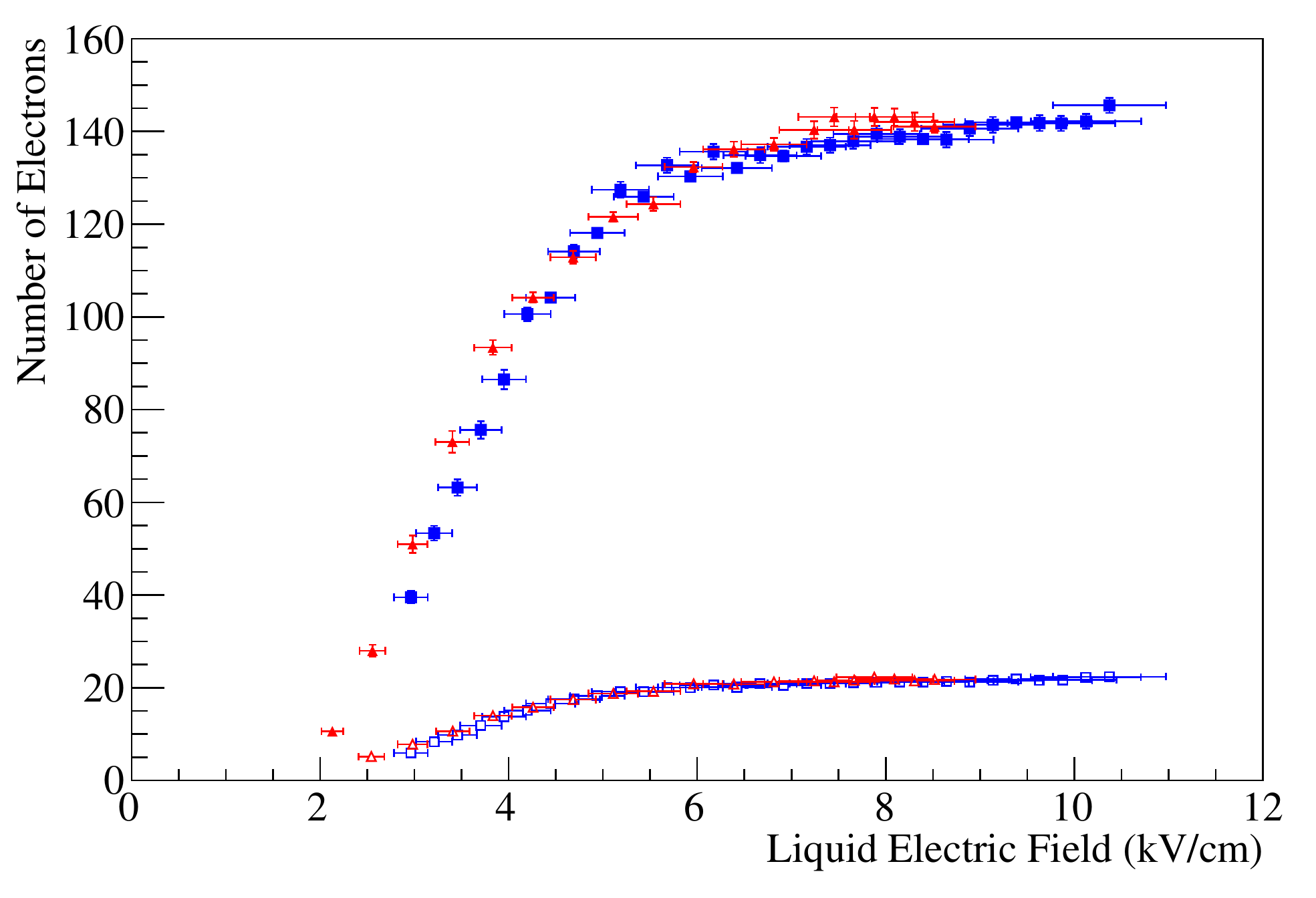}
\caption{The calculated numbers of electrons from \arsev\ decays as a function of extraction electric field: 
LF K-shell (solid triangles, red), 
LF L-shell (hollow triangles, red), 
HF K-shell (solid squares, blue), 
and HF L-shell(hollow squares, blue). 
The electric field uncertainties mostly originate from the different grid configurations between the LF and HF measurements, 
and the electron number uncertainties ($\sim$1-3\%) include that from the SE fits and the \arsev\ peak fits. 
}
\label{fig:arne}
\end{figure}

The numbers of detected electrons for both K-shell and L-shell \arsev\ decays at different extraction electric fields,
calculated using SE calibrations from Figure~\ref{fig:sevse}, 
are shown in Figure~\ref{fig:arne}. 
For L-shell \arsev\ decays, 
the peaks were simply fitted with a single Gaussian function on top of an exponential background, 
and the peak positions did not exhibit significant difference for decays above and below the EG. 
The average numbers of electrons for the K-shell and L-shell \arsev\ decays at $\sim$7.1~kV/cm extraction field measured in this experiment 
are 138.4$\pm$2.9 and 21.2$\pm$0.5, respectively, 
consistent with that of 134.7$\pm$8.9 and 20.9$\pm$1.3 by the PIXeY experiment at similar drift and extraction field values~\cite{Boulton2017_Ar37}. 
The slight shift between the LF data and the HF data that is observed in both the K-shell result and the L-shell result 
can be largely attributed to systematic uncertainties in the estimated drift/extraction field values in the two different hardware configurations, 
as indicted by the horizontal error bars (including both systematic and statistical uncertainties). 
In addition, the purity levels of liquid xenon in these two measurements may differ slightly, 
leading to different losses of electrons during the drifting process before they can be extracted. 
Using the scintillation-EL coincidence for K-shell \arsev\ decay events, 
we calculated the electron lifetime values to estimate the electron loss to impurities, 
and introduced an additional systematic uncertainty of 1\% to account for the impurity variation. 

\section{Discussions}
\label{sec:result}

\begin{figure}[h!]
\centering
\includegraphics[width=.42\textwidth]{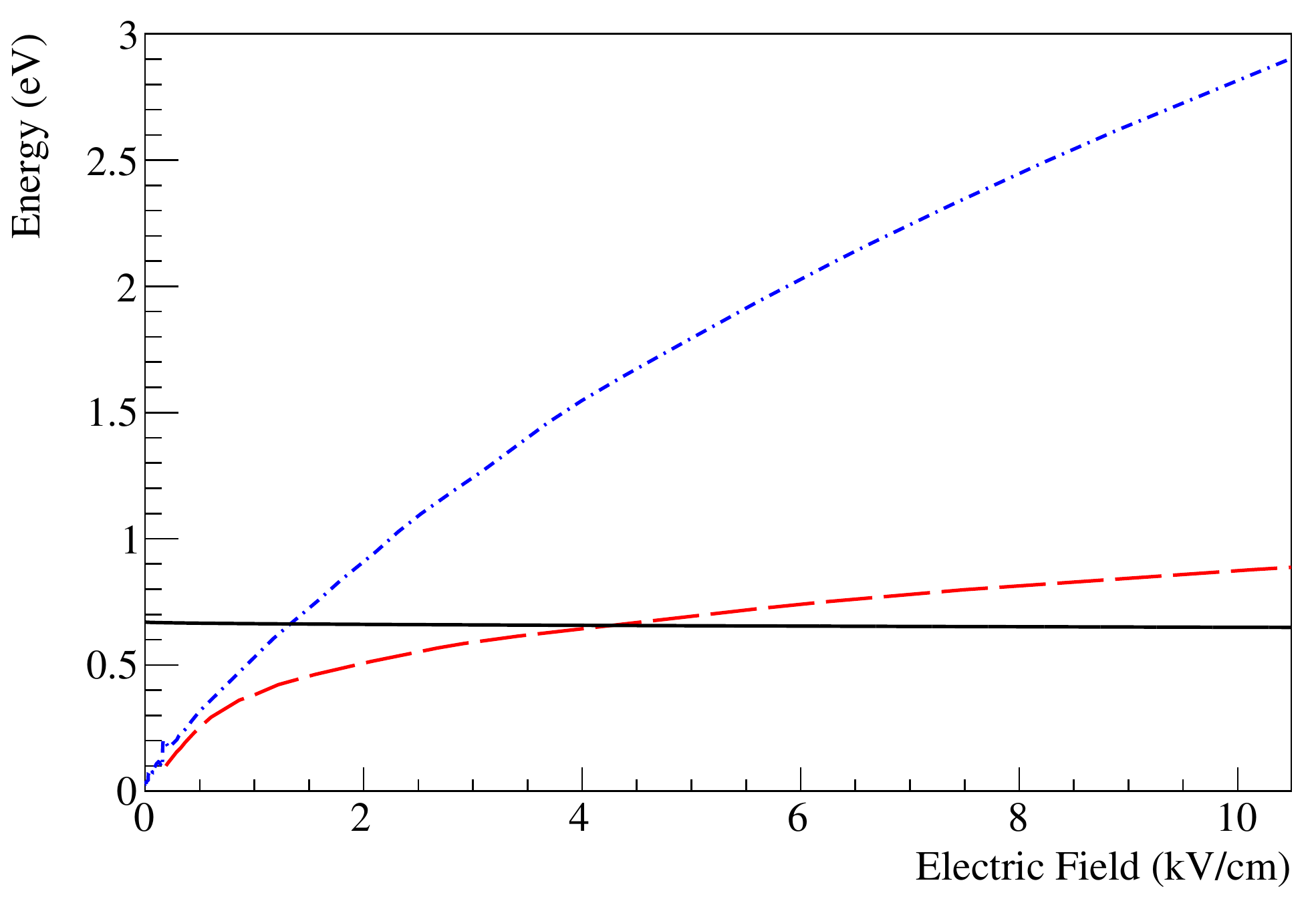}
\caption{the calculated mean electron energy in liquid xenon as a function of the extraction electric field under two transport models 
(red dashed curve~\cite{Atrazhev1985_EHeating} and blue dash-dot curve~\cite{Gushchin1982_ETransport}), 
in comparison with the estimated potential barrier height at the liquid surface~\cite{Bolozdynya1995_XeTPC, Gushchin1982_EEE}. 
The barrier height at zero electric field is assumed to be 0.67~eV for this illustration. 
}
\label{fig:eke}
\end{figure}

In Figure~\ref{fig:arne}, 
the number of detected electrons in the gas phase 
increases rapidly with the applied extraction electric field below 6~kV/cm for both K-shell and L-shell \arsev\ decays. 
At higher field values, 
the dependence becomes very mild. 
Particularly, for field strengths of $>$7.5~kV/cm, 
the increase in the collected electron number with the extraction field is at the level of $<$1\%(kV/cm)$^{-1}$. 
This value is smaller than the measurement uncertainty at each data point, 
suggesting that a saturation of the EEE may have finally been observed. 
As discussed in Section~\ref{sec:mechanism}, 
the number of detected electrons from mono-energetic sources, 
which is proportional to the EEE at the liquid surface, 
would keep increasing with the electron energy until 100\% EEE is achieved. 
The average energy of drifting electrons in liquid xenon, 
as calcualted from electron transport models~\cite{Atrazhev1985_EHeating, Gushchin1982_ETransport}, 
is expected to increase monotonically with the applied electric field over the field range studied in this experiment,  
as illustrated in Figure~\ref{fig:eke}. 
Therefore, the saturation features in Figure~\ref{fig:arne} serve as a strong indication that 
close to 100\% absolute EEE value has been achieved in this measurement, 
despite the measurement being relative in nature. 

\begin{figure}[h!]
\centering
\includegraphics[width=.45\textwidth]{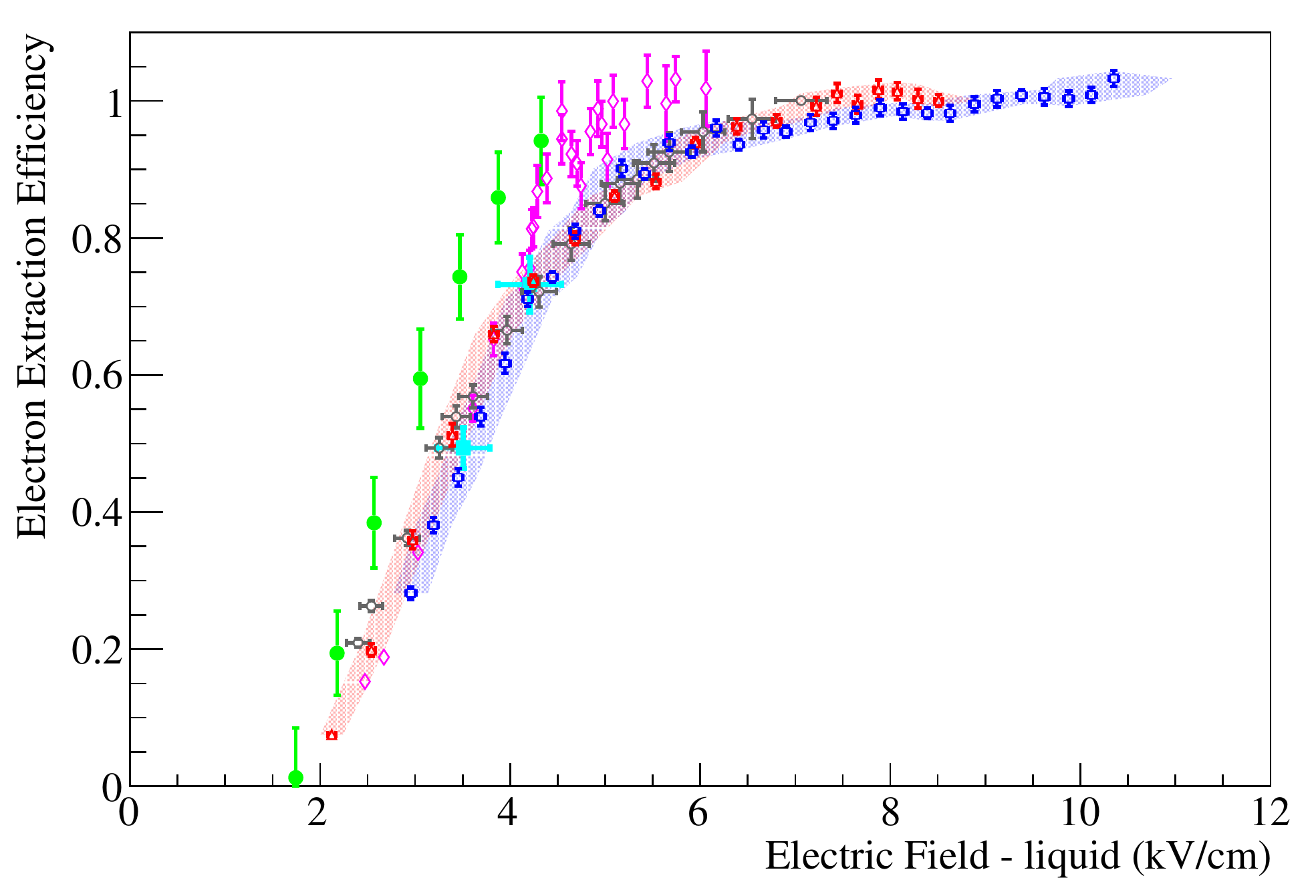}
\caption{The derived EEE values as a function of extraction electric field (in liquid xenon) 
for both the LF run (red triangles) 
and the HF run (blue squares). 
100\% EEE is defined as the average value between 7.5 and 10.4~kV/cm. 
Systematic uncertainties in the measurements are illustrated using the bands. 
For comparison, we also show the relative EEE measurement results from 
XENON100~\cite{XENON2014_SE} (magenta diamonds) 
and PIXeY~\cite{PIXeY2018_EEE} (grey circles), 
along with the absolute EEE results from Gushchin {\it et al.}~\cite{Gushchin1979_EEE} (green dots), 
and LUX~\cite{LUX2016_Run3_4} (cyan squares, extraction field values are calculated using the reported geometries and the assumption of 0.5$\pm$0.5~mm higher liquid level than the spillover resevior due to fluid dynamics).  
Relative measurements are shown as hollow markers, 
while absolute measurements are shown as solid markers. }
\label{fig:eee}
\end{figure}

In the discussions hereafter, we assume the average EEE value in the field range of 7.5-10.4~kV/cm to be 100\%, 
and then the EEE values at different extraction electric fields can be calcualted through a simple scaling. 
Thanks to the excellent agreement between the K-shell result and the L-shell result for both LF and HF measurements,  
we summed the numbers of detected electrons from the two decay modes for higher accuracy, 
and the resulted EEE curve is shown in Figure~\ref{fig:eee}. 
We comment that the slight increase of the EEE above 7.5~kV/cm may be explained by some systematic effects that are not accounted for in this analysis. 
For example, 
the electron transparency of the EG and the electric field leakage from above the EG to below 
may both increase slightly with the extraction electric field, 
causing more electrons to be observed in the measurements. 
Alternatively, it could be a result of the low-energy tail of the electron energy distribution being slowly lifted 
to above the potential barrier at the liquid surface. 
In this scenario, 
100\% absolute EEE may only be achieved asymptotically at higher electric field values. 
Due to the falling tail of the heated electron energy distribution 
we expect the EEE slope at higher fields to be smaller than $\sim$1\%/(kV/cm) if this hypothesis were true. 
Therefore, the absolute 100\% EEE--if different from the highest EEE value measured in this work--should not deviate 
from this result by more than a few percent. 
A measurement at even stronger extraction field, 
which may be achievable with an upgrade to this experimental setup, 
can further clarify the situation. 

Figure~\ref{fig:eee} also compares this EEE result with other measurements in literature. 
Of all of these measurements, 
this work covers the largest electric field range of up to 10.4~kV/cm in the liquid, 
compared to $<$7.1~kV/cm in PIXeY~\cite{PIXeY2018_EEE}, $<$6.1~kV/cm in XENON100~\cite{XENON2014_SE}, 
and $<$4.3~kV/cm in Gushchin {\it et al.}~\cite{Gushchin1979_EEE}. 
This result agrees with that of PIXeY, XENON100, and LUX~\cite{LUX2016_Run3_4} (indirect method, Equation~\ref{eq:corr}) at low field values; 
at high fields, 
the discrepancies with XENON100 and PIXeY are likely to be the result of the different scaling factors used when the experimental results are reported. 
In the relative EEE scale used in this work, 
the highest EEE value measured in XENON100 corresponds to $\sim$92\% efficiency, 
and that in PIXeY corresponds to $\sim$96\% efficiency. 
The Gushchin experiment was designed to measure the absolute EEE values, 
so it is not subject to such biases; 
but due to the lack of details in \cite{Gushchin1979_EEE} 
we do not attempt to resolve the discrepancy. 

The highest electric field covered in this experiment far exceeds that used in any existing 
or proposed xenon-based dark matter experiments. 
With the observation of an apparent EEE saturation and the excellent agreement with other recent measurements, 
this work offers the most comprehensive calibration of EEE for dual-phase xenon TPC experiments to date. 
Using the EEE scale in Figure~\ref{fig:eee}, 
$\sim$10-15\% of ionization electrons were left un-extracted in the XENON10 and XENON100 experiments, 
to the contrary of the assumed $\sim$100\% electron extraction; 
re-emission of these electrons can possibly explain the high observed background rates in the charge-only dark matter searches~\cite{Essig2012_SubGeVXENON10, Essig2017_SubGeVXENON100}. 
Characterization and reduction of this unextracted electron background will help us achieve a complete understanding of the 
low-energy ionization-like background observed in xenon-based dark matter experiments~\cite{ZEPLIN2011_SE, XENON2014_SE, LUX2016_SE, DS50_2018_S2Only}. 
If a substantially lower background electron level can be achieved, 
a compact detector at the order of $\sim$10~kg may offer compelling sensitivity to 
low-mass WIMPs and certain dark sector dark matter particles~\cite{CosmicVision2017_WP}. 
In addition, such detector development may also enable the monitoring of reactor anti-neutrinos using compact noble liquid TPCs~\cite{Hagmann2004_CENNS} 
through the recently demonstrated coherent elastic neutrino nucleus scattering process~\cite{COHERENT2017_CENNS}. 

Lastly, this experimental result on the extraction of hot electrons from liquid xenon into the gas phase under the influence of electric field 
also contributes to the studies of hot electron transport in non-polar liquid 
and across phase boundaries. 
Noble liquids such as xenon resemble the simplest dense matter and disordered systems, 
electron dynamics in which have been the topic of continuing studies in condensed matter physics~\cite{Frost1964_ETransport, CohenLekner1967_ETransport, Atrazhev1981_HotE, Gushchin1982_ETransport, Atrazhev1985_EHeating, Boyle2016_ETransport, Gordon2001_ETransport, Sakai1984_ETransport}, 
plasma physics~\cite{Garland2018_HotEInterface}, and laser developments~\cite{Schussler2000_LXeEL}. 

\section{Conclusion}
\label{sec:concl}

We report a new measurement of the efficiency of extracting electrons from liquid xenon into gas 
over a large range of extraction electric field, 
which is a key performance parameter for xenon-based dark matter experiments. 
By demonstrating previously unattained high voltage performance, 
we studied the EEE values at the highest electric field strength reported to date. 
For the first time, 
a strong evidence of EEE saturation is observed over a large electric field window of 7.5-10.4~kV/cm. 
Combining this observation with electron transport and emission models developed for liquid xenon, 
we suggest that this relative EEE result may be used to infer the absolute EEE scale. 
This result offers the most comprehensive electron extraction efficiency calibration for both existing and future xenon TPC experiments. 
It also provides valuable information for xenon-based experiments to obtain a better understanding of their low electron background 
and thus improve their potential sensitivity to low-energy dark matter interactions 
and to reactor antineutrinos. 

\begin{acknowledgments}

 This project is supported by the U.S. Department of Energy (DOE) Office of Science, 
 Office of High Energy Physics under Work Proposal Number SCW1508 and SCW1077 awarded to Lawrence Livermore National Laboratory (LLNL). 
 LLNL is operated by Lawrence Livermore National Security, LLC, for the DOE, National Nuclear Security Administration (NNSA) under Contract DE-AC52-07NA27344.
 Certain equipment used in this measurement was recycled from the LLNL DUS Second Time Around store.  
 D.~Naim is supported by the DOE/NNSA under Award Number DE-NA0000979 through the Nuclear Science and Security Consortium. 
 
 We thank Sean Durham, Jesse Hamblen from LLNL, 
 and Dave Hemer, Keith Delong and Michael Irving from UC Davis for their technical support 
 on designing and constructing the testbed xenon TPC detector. 
 We thank the staff at McClellan Nuclear Research Center for working with us to generate the $^{37}$Ar source used in this work.
 We thank the LUX/LZ collaboration for their constructive discussions on this measurement and possible implications. 

\end{acknowledgments}

\bibliographystyle{apsrev}
\bibliography{biblio}

\end{document}